\documentstyle[12pt,aps,prd,preprint]{revtex}
\begin{document}
\title{Late-Time Evolution of Realistic Rotating Collapse and The
  No-Hair Theorem}
\author{Shahar Hod}
\address{The Racah Institute for Physics, The
Hebrew University, Jerusalem 91904, Israel}
\date{\today}
\maketitle

\begin{abstract}
We study {\it analytically} the asymptotic late-time evolution of 
realistic {\it rotating} collapse. This is done by considering the
asymptotic late-time solutions of Teukolsky's master equation, 
which governs the evolution 
of gravitational, electromagnetic, neutrino and scalar 
perturbations fields on Kerr spacetimes.
In accordance with the {\it no-hair conjecture} for rotating
black-holes we show that the asymptotic solutions develop 
inverse {\it power-law} tails at the asymptotic regions of 
timelike infinity, null infinity and along the black-hole 
outer horizon (where the power-law behaviour is multiplied 
by an oscillatory term caused by the dragging of reference frames).
The damping exponents characterizing the asymptotic solutions 
at timelike infinity and
along the black-hole outer horizon are independent of the spin
parameter of the fields. However, the damping exponents 
at future null infinity are {\it spin dependent}.
The late-time tails at all the three asymptotic regions 
are spatially dependent on the spin parameter of the field.
The {\it rotational} dragging of reference frames, 
caused by the rotation of the black-hole (or star) 
leads to an active {\it coupling} of different multipoles.
\end{abstract}

\section{introduction}\label{Sec1}

The {\it no-hair theorem}, introduced by Wheeler in the
early 1970s \cite{Whee}, states that the external
field of a black-hole relaxes to a Kerr-Newman field characterized solely
by the black-hole's mass, charge and angular-momentum. 
The mechanism for this relaxation process of neutral fields was first analyzed 
by Price \cite{Price} for a nearly spherical collapse. 
The physical mechanism by which 
a charged black-hole, which is formed during a 
gravitational collapse of a {\it charged} matter, dynamically 
sheds its charged hair was first studied in \cite{HodPir1,HodPir2}.
However, these analysis were restricted to spherically
symmetric backgrounds, i.e., to the Schwarschild and 
Reissner-Nordstr\"om black-holes.
On the other hand, the physical process of stellar core 
collapse to form a black-hole is expected
to be {\it non-}spherical in nature because of stellar rotation. The
analogous problem of the dynamics of massless perturbations outside
realistic, {\it rotating} black-holes is much more complicated due to
the lack of spherical symmetry. 

Recently, this problem was addressed
numerically \cite{Krivan1,Krivan2} and analytically for the case of
scalar-perturbations \cite{Ori}.
However, the evolution of higher-spin fields (and in
particular {\it gravitational} perturbations) outside realistic, 
rotating black-holes was not analyzed analytically so far.
In addition, the works done so far were restricted to the
asymptotic regions of timelike infinity \cite{Krivan1,Krivan2,Ori} and
along the black-hole outer-horizon \cite{Ori}. The asymptotics at 
null infinity of rotating collapse was not studied so far.
In this paper we give for the first time 
a full analytic analysis of the late-time evolution of 
realistic rotating collapse. Our analysis considers massless fields with
arbitrary spin (and in particular {\it gravitational} perturbations).
In addition, we consider the late-time evolution at {\it all} 
the three asymptotic regions: timelike infinity, future null infinity
and along the black-hole outer horizon.

There are two different approaches to the study of perturbations of
Kerr spacetimes. The first is to consider perturbations of the metric
functions. However, this direct approach leads to gauge-dependent
formulations. An alternative approach is to consider {\it curvature}
perturbations (perturbations of the Weyl scalars).
Based on the tetrad formalism by Newman and Penrose, Teukolsky derived
a master equation governing the perturbations of Kerr spacetimes 
\cite{Teukolsky1,Teukolsky2}. The main goal of this paper is to
analyze the asymptotic {\it late}-time evolution of a realistic
{\it rotating} collapse. This is done by studying the asymptotic late-time 
solutions of the Teukolsky equation.

The plan of the paper is as follows. In Sec. \ref{Sec2} we give a
short description of the physical system and formulate the evolution
equation considered. In Sec. \ref{Sec3} we formulate the problem in
terms of the black-hole Green's function using the technique of
spectral decomposition.  In Sec. \ref{Sec4} we study the late-time
evolution of perturbations of Kerr spacetimes, i.e., the
asymptotic late-time solutions of the Teukolsky equation.
We conclude in Sec. \ref{Sec5} with a brief summary of our results.

\section{Description of the system}\label{Sec2}

We consider the evolution of gravitational, electromagnetic, neutrino
and scalar (massless) perturbations 
fields outside a {\it rotating} collapsing star.
The external gravitational field of a rotating object of
mass $M$ and angular-momentum per unit-mass $a$ is given by the
Kerr metric, which in Boyer-Lindquist coordinates takes the form
\begin{eqnarray}\label{Eq1}
ds^{2}&=& \left( 1 -{{2Mr} \over \Sigma}\right) dt^{2} +
\left( {{4Marsin^{2} \theta} \over \Sigma} \right) dtd \varphi-
{\Sigma \over \Delta} dr^{2} -\Sigma d\theta^{2} - \nonumber \\
&& sin^{2} \theta \left( r^{2} +a^{2} +{{2Ma^{2}rsin^{2} \theta} \over
    \Sigma} \right) d\varphi^{2} \  ,
\end{eqnarray}
where $\Sigma=r^{2}+a^{2}cos^{2} \theta$ and $\Delta=r^{2} -2Mr
+a^{2}$. Throughout this paper we use $G=c=1$.

Using the Newman-Penrose formalism, Teukolsky
\cite{Teukolsky1,Teukolsky2} derived a master equation that governs 
the evolution of perturbations of the Kerr spacetime
\begin{eqnarray}\label{Eq2}
&& \left[{{(r^{2}+a^{2})^{2}} \over {\Delta}} 
-a^{2}sin^{2}\theta \right]
{{\partial ^2 \psi} \over {\partial t^2}} 
+{{4Mar} \over {\Delta}} 
{{\partial ^2 \psi} \over {\partial t \partial \varphi}} 
+ \left[{{a^{2}} \over \Delta} -{1 
\over {sin^{2} \theta}} \right] {{\partial ^2 \psi} 
\over {\partial \varphi ^2}} \nonumber \\
&& -\Delta^{-s} {\partial \over {\partial r}} 
\left( \Delta^{s+1} {{\partial \psi} 
\over {\partial r}} \right) -{1 \over {sin\theta}} 
{{\partial} \over {\partial
    \theta}} \left( sin \theta{{\partial \psi} \over {\partial
    \theta}} \right) -2s \left[{{a(r-m)} \over \Delta} 
+{{icos \theta} \over {sin^{2} \theta
}} \right] {{\partial \psi} \over {\partial
    \varphi}} \nonumber \\
&& -2s \left[{{M(r^{2}-a^{2})} \over \Delta} -r -iacos 
\theta \right] {{\partial \psi} \over
 {\partial t}} +(s^{2}cot^{2} \theta -s) \psi =0\ \ ,
\end{eqnarray}
where the parameter $s$ is the spin-weight of the field.
The Teukolsky's master equation is valid for scalar perturbations fields
($s=0$), neutrino perturbations fields ($s= \pm 1/2$), electromagnetic
perturbations fields ($s= \pm 1$) and gravitational 
perturbations ($s= \pm 2$).
The field quantities $\psi$ which satisfy this equation (for the 
various values of the spin parameter $s$) are given in \cite{Teukolsky2}.
Resolving the field in the form
\begin{equation}\label{Eq3}
\psi= \Delta^{-s/2} (r^{2}+a^{2})^{-1/2} 
\sum\limits_{m= -\infty}^{\infty} 
\Psi^{m} e^{im \varphi}\ \ ,
\end{equation}
one obtains a wave-equation for each value of $m$
\begin{equation}\label{Eq4}
B_{1}(r,\theta) {{\partial ^2 \Psi} \over {\partial t^2}} +
B_{2}(r,\theta) {{\partial \Psi} \over
{\partial t}} -
{{\partial ^2 \Psi} \over {\partial y^2}}+ B_{3}(r,\theta) \Psi - 
{{\Delta} \over {(r^{2}+a^{2})^2}} {1 \over {sin\theta}} 
{{\partial} \over {\partial
    \theta}} \left( sin \theta{{\partial \Psi} \over {\partial
    \theta}} \right)=0\ \ ,
\end{equation}
where the tortoise radial coordinate $y$ is defined 
by $dy={{r^{2}+a^{2}} \over \Delta} dr$.
The coefficients $B_{i}(r,\theta)$ are given by
\begin{equation}\label{Eq5}
B_{1}(r,\theta)=1-{{\Delta a^{2}sin^{2}\theta} \over {(r^{2}+a^{2})^{2}}}\ \ ,
\end{equation}
\begin{equation}\label{Eq6}
B_{2}(r,\theta)=\Bigg \{ {{4iMmar} \over \Delta} -2s \left[ {{M(r^{2}-a^{2})} \over \Delta} 
-r -iacos \theta \right] \Bigg \} {\Delta \over {(r^2+a^2)^2}}\ \ ,
\end{equation}
and
\begin{eqnarray}\label{Eq7}
B_{3}(r,\theta)&=&\Bigg \{ 2(s+1)(r-M) \left[ s(r-M) \Delta^{-1} +
r(r^2+a^2)^{-1} \right] -m^{2} \left[ {{a^{2}} 
\over \Delta} -{1 \over {sin^{2}\theta}} \right] - \nonumber \\
&& 2sim \left[ {{a(r-M)} \over \Delta} + {{icos\theta} 
\over {sin^{2}\theta}} \right] +(s^{2
}cot^{2}\theta -s) \Bigg \} {\Delta \over {(r^2+a^2)^2}}\ \ .
\end{eqnarray}

\section{Formalism}\label{Sec3}

The time-evolution of a wave-field 
described by Eq. (\ref{Eq4}) is given by
\begin{eqnarray}\label{Eq8}
\Psi (z,t) &=& 2\pi \int \int_{0}^{\pi} \Bigg\{ B_{1}(z') 
\Big [ G(z,z';t) \Psi _t(z',0)+G_t(z,z';t) 
\Psi (z',0) \Big] +  \nonumber \\ 
&& B_{2}(z') G(z,z';t) \Psi (z',0) \Bigg\} sin\theta' d\theta'dy'\ \ ,
\end{eqnarray}
for $t>0$, where $z$ stands for $(y,\theta)$.
The (retarded) Green's function $G(z,z';t)$ is defined by

\begin{eqnarray}\label{Eq9}
&&\left [ B_{1}(r,\theta) {{\partial ^2} \over {\partial t^2}} +
B_{2}(r,\theta) {{\partial} \over
{\partial t}} -
{{\partial ^2} \over {\partial y^2}}+ B_{3}(r,\theta) - 
{{\Delta} \over {(r^{2}+a^{2})^2}} {1 \over {sin\theta}} 
{{\partial} \over {\partial
    \theta}} \left( sin \theta{{\partial} \over {\partial
    \theta}} \right) \right ] G(z,z';t)=  \nonumber \\
&&  \delta (t) \delta(y-y') 
{{\delta(\theta - \theta')} \over {2\pi sin\theta}} \ \ .
\end{eqnarray}
The causality condition gives us the initial condition $G(z,z';t)=0$ for
$t \leq 0$.
In order to find $G(z,z';t)$ we use the Fourier transform 
\begin{equation}\label{Eq10}
\tilde G_{l}(y,y';w)_{s}S_{l}^{m}(\theta',aw) =2\pi \int_{0^-}^{\infty} \int_{0}^{\pi} 
G(z,z';t) _{s}S_{l}^{m}(\theta,aw) sin\theta e^{iwt} d\theta dt\ \ ,
\end{equation}
where $_{s}S_{l}^{m}(\theta,aw)$ are the spin-weighted spheroidal
harmonics which are solutions to the angular-equation \cite{Teukolsky2}
\begin{eqnarray}\label{Eq11}
&& {1 \over {sin\theta}} 
{{\partial} \over {\partial
    \theta}} \left( sin \theta{{\partial} \over {\partial
    \theta}} \right) + \nonumber \\
&&  \left(a^{2}w^{2}cos^{2}\theta-{{m^{2}} \over
  {sin^{2}\theta}} -2awscos \theta - {{2mscos\theta} \over
  {sin^{2}\theta}} - s^{2}cot^{2}\theta + s +{_{s}A_{l}^{m}} \right) 
{ _{s}S_{l}^{m}}  =0  \  .
\end{eqnarray}
For the $aw=0$ case, the eigenfunctions $_{s}S_{l}^{m}(\theta,aw)$
reduce to the spin-weighted spherical harmonics
$_{s}Y_{l}^{m}(\theta,\phi)={_{s}S_{l}^{m}(\theta)e^{im\varphi}}$, and the
separation constants $_{s}A_{l}^{m}(aw)$ are 
simply $_{s}A_{l}^{m}=(l-s)(l+s+1)$ \cite{Goldberg}. 

The Fourier transform is analytic in the upper half $w$-plane and it satisfies
the equation \cite{Teukolsky2}
\begin{equation}\label{Eq12}
\Bigg \{{{d^2} \over {dy^2}} + \left[ {{K^{2} -2is(r-M)K+
    \Delta(4irws-\lambda)} \over {(r^{2}+a^{2})^{2}}}-H^{2}-{{dH} \over
    {dy}} \right] \Bigg\} \tilde G_{l}(y,y';w)= \delta(y-y')  \  ,
\end{equation}
where $K=(r^{2}+a^{2})w-am$, $\lambda=A+a^{2}w^{2}-2amw$ and 
$H=s(r-M)/(r^{2}+a^{2})+r\Delta/(r^{2}+a^{2})^{2}$.
$G(z,z';t)$ itself is given by the inversion formula
\begin{equation}\label{Eq13}
G(z,z';t)={1 \over {(2 \pi)^{2}}} \sum\limits_{l=l_{0}}^{\infty} 
\int_{- \infty +ic}^{\infty +ic}
\tilde G_{l}(y,y';w)_{s}S_{l}^{m}(\theta,aw)_{s}S_{l}^{m}(\theta',aw)
e^{-iwt} dw\  ,
\end{equation}
where $c$ is some positive constant and $l_{0}=max(|m|,|s|)$.

Next, we define two auxiliary functions $\tilde \Psi_1(z,w)$ and 
$\tilde \Psi_2(z,w)$ which are (linearly independent) solutions to the 
homogeneous equation
\begin{equation}\label{Eq14}
\Bigg[ {{d^2} \over {dy^2}} +  {{K^{2} -2is(r-M)K+
    \Delta(4irws-\lambda)} \over {(r^{2}+a^{2})^{2}}}-H^{2}-{{dH} \over
    {dy}} \Bigg] \tilde \Psi_{i}(y,w)=0  \ ,  \ \ \  i=1,2  \ .
\end{equation}
The two basic solutions that are required in order to build the black-hole
Green's function are defined by their asymptotic behaviour:
\begin{equation}\label{Eq15}
\tilde \Psi_1(y,w) \sim 
\left\{
 \begin{array}{l@{\quad,\quad}l}
\Delta^{-s/2} e^{-iky} & y \to -\infty\  , \\
A_{out}(w) y^{i2wM-s} e^{iwy} + A_{in}(w) y^{-i2wM+s} e^{-iwy} & y \to \
\infty\  ,
\end{array} \right. \    
\end{equation}
and
\begin{equation}\label{Eq16}
\tilde \Psi_2(y,w) \sim 
\left\{
 \begin{array}{l@{\quad,\quad}l}
B_{out}(w) \Delta^{s/2} e^{iky} +B_{in}(w) \Delta^{-s/2}
e^{-iky} & y \to -\infty\  , \\
y^{i2wM-s} e^{iwy} & y \to \infty\  ,
\end{array} \right.
\end{equation}
where $k=w-mw_{+}$, $w_{+}=a/(2Mr_{+})$.
Let the Wronskian be
\begin{equation}\label{Eq17}
W(w)=W(\tilde \Psi_1, \tilde \Psi_2 )= \tilde \Psi_1 \tilde \Psi_{2,y} - 
\tilde \Psi_2 \tilde \Psi_{1,y}\  ,
\end{equation}
where $W(w)$ is $y$-independent.
Thus, using the two solutions $\tilde \Psi_1$ and $\tilde \Psi_2$, the
black-hole Green's function can be expressed as
\begin{equation}\label{Eq18}
\tilde G_{l}(y,y';w) =- {1 \over {W(w)}} 
\left\{ \begin{array}{l@{\quad,\quad}l}
\tilde \Psi_1(y,w) \tilde \Psi_2(y',w) & y<y' \  , \\
\tilde \Psi_1(y',w) \tilde \Psi_2(y,w) & y>y' \  .
\end{array} \right.
\end{equation}

In order to calculate $G(z,z';t)$ using Eq. (\ref{Eq13}), one may close the 
contour of integration into the lower half of the complex frequency plane.
Then, one finds three distinct contributions to $G(z,z';t)$ \cite{Leaver} :
\begin{enumerate}
\item {\it Prompt contribution}. This comes from the integral along the large
semi-circle. This term contributes to the {\it short}-time 
response.

\item {\it Quasinormal modes}. This comes from the distinct singularities of
$\tilde G(y,y';w)$ in the lower half of the complex $w$-plane, and it 
is just the sum of the residues at the 
poles of $\tilde G(y,y';w)$. 
Since each mode has Im$w<0$ this term decays 
{\it exponentially} with time. 
\item {\it Tail contribution}. The late-time tail is
associated with the existence of a branch cut (in $\tilde \Psi_2$) \cite{Leaver}
, usually
placed along the negative imaginary $w$-axis. This tail arises from the 
integral of $\tilde G(y,y';w)$ around the branch cut and is denoted by  $G^C$.
As will be shown, the contribution $G^C$ leads to an inverse {\it power-law}
behaviour (multiplied by a periodic term along the black-hole outer horizon) of
the field. Thus, $G^C$ dominates the late-time behaviour of the field.
\end{enumerate}

The present paper focuses on the late-time asymptotic solutions of the 
Teukolsky equation. Thus, the purpose of this paper is to 
evaluate $G^C(z,z';t)$.

\section{Late-Time Behaviour of Realistic Rotating Collapse}\label{Sec4}

\subsection{The large-$r$ (low-$w$) approximation}\label{Sec4A}

It is well known that the late-time behaviour of massless perturbations 
fields is determined by the backscattering from asymptotically {\it far}
regions \cite{Thorne,Price}. Thus, the late-time behaviour is dominated by the
{\it low}-frequencies contribution to the Green's function, for only low
frequencies will be backscattered by the small effective 
potential (for $r \gg M$) in Eq. (\ref{Eq14}). 
Thus, as long as the observer is situated far away from the 
black-hole and the initial data has a considerable support only far away from 
the black-hole, a {\it large}-$r$ (or equivalently, a 
{\it low}-$w$) approximation is sufficient in order to study the asymptotic {\it
 late-time}
behaviour of the field \cite{Andersson}.
Expanding Eq. (\ref{Eq14}) for large $r$ one obtains 
[using $\lambda=(l-s)(l+s+1)+O(aw)$ and neglecting terms
of order $O({w \over {r^2}})$ and higher]
\begin{equation}\label{Eq19}
\left[ {{d^2} \over {dr^2}} +w^{2} +{{4Mw^{2}+2isw} \over r} - 
{{l(l+1)} \over {r^{2}}} \right ] \tilde \Psi =0\  .
\end{equation}
It should be noted that this equation is the {\it correct} form 
of equation (66) of Ref. \cite{Andersson} (the term $s^2 / r^2$ should
{\it not} appear in that equation). 

We introduce a second auxiliary field $\tilde \phi$ defined by
\begin{equation}\label{Eq20}
\tilde \Psi =r^{l +1} e^{iwr} \tilde \phi (x)\  ,
\end{equation}
where $x=-2iwr$.
$\tilde \phi(x)$ satisfies the confluent hypergeometric equation
\begin{equation}\label{Eq21}
\left[ x {{d^{2}} \over {dx^{2}}} +(2l+2-x) {d \over {dx}} -(l+1-
2iw\alpha) \right] \tilde \phi(x)=0\  ,
\end{equation}
where
\begin{equation}\label{Eq22}
\alpha=M+{{is} \over {2w}}\  .
\end{equation}
Thus, the two basic solutions required in order to build 
the Green's function are (for $r \gg M,|a|$)
\begin{equation}\label{Eq23}
\tilde \Psi_1 =Ar^{l +1} e^{iwr} M(l+s+1-2iwM ,2l +2, 
-2iwr)\  ,
\end{equation}
and
\begin{equation}\label{Eq24}
\tilde \Psi_2 =Br^{l +1} e^{iwr} U(l+s+1-2iwM ,2l +2, 
-2iwr)\  ,
\end{equation}
where $A$ and $B$ are normalization constants. 
$M(a,b,z)$ and $U(a,b,z)$ are 
the two standard solutions to the confluent hypergeometric equation 
\cite{Abram}.
From these solutions it is clear that the black-hole's rotation
parameter $a$ is irrelevant in
the context of the {\it asymptotic} form of the Green's function (to
leading order in the inverse time). However, as will be shown below, 
the {\it rotational} dragging of reference frames, 
caused by the rotation of the black-hole (or star), has (two) important 
effects (which are absent in the non-rotating case) on
the asymptotic late-time evolution of {\it rotating} collapse.
Hence, in order to find the asymptotic form of the Green's function 
it is sufficient to analyze the late-time solutions in
the {\it spherically} symmetric limit (i.e, on the Schwarchild background).

A note is needed here on the relations between the various
equations governing the perturbations of the Schwarchild spacetime.
There are two different approaches to the study of linearized gravitational
perturbations: by considering metric perturbations, or, 
alternatively, by considering perturbations of the Weyl scalars.
Equations governing metric perturbations of the Schwarchild 
spacetime were derived by Regge and Wheeler \cite{RegWheel} 
(for odd-parity perturbations) 
and by Zerilli \cite{Zer} (for polar perturbations).
An alternative approach, based on the tetrad formalism by Newman 
and Penrose was used by Teukolsky [7,8] to derive a separable 
wave equation for perturbations of the Weyl Scalars. These 
scalars, constructed from the Weyl tensor in a given
tetrad basis, characterize the gravitational field in 
vacuum. They allow a more convenient approach to the study of 
gravitational perturbations, due to their scalar nature.
As was first shown by Chandrasekhar \cite{Chan}, the $a \to 0$ limit
of the Teukolsky equation, the Bardeen-Press equation \cite{BarPres},
can easily be transformed into the Regge-Wheeler equation. However,
this equivalence is restricted to the {\it frequency} domain. In other
wards, any solution (in the frequency domain) of the Bardeen-Press 
equation can be transformed
in a trivial manner into a solution of the Regge-Wheeler equation, and
vise versa. However, these solutions have, of coarse, a {\it different}
$w$-dependence (as can be verified for example from their 
asymptotic forms Eqs. (\ref{Eq23}) and
(\ref{Eq24})). Hence, this difference in the $w$-dependence of the 
solutions, when integrated in the complex frequency plane (in
order to obtain the {\it temporal} dependence of the solutions) 
may lead to a non-trivial difference in the time dependence of
the solutions. Thus, it is not apriori guaranteed that 
the asymptotic late-time solutions of 
Teukolsky's equation and the Regge-Wheeler equation (studied 
by price \cite{Price} for the spherically symmetric case)
will have the same temporal behaviour
(Indeed, it is found that the damping exponents, describing the
fall-off of the asymptotic solutions at future null infinity are 
spin-dependent, a phenomena not found for the Regge-Wheeler asymptotic
solutions, see Sec. \ref{Sec4C} below).
Furthermore, one can easily verify that even the $a \to 0$ limit
of the Teukolsky equation depends on the spin parameter $s$ [see
Eq. (\ref{Eq2}) and its asymptotic form Eq. (\ref{Eq19})]. 
Hence, it is not surprising that the asymptotic late-time solutions of the
Teukolsky equation will turn out to be spin-dependent.

The function $U(a,b,z)$ is a many-valued function, i.e., there will be
a cut in $\tilde \Psi_2$.
Using Eq. (\ref{Eq13}), one finds that the branch cut contribution to the 
Green's function is given by
\begin{eqnarray}\label{Eq25}
G^C(z,z';t)& =&  {1 \over {(2\pi)^2}} \sum\limits_{l=l_{0}}^{\infty}  
\int_{0}^{-i \infty} 
\tilde \Psi_1(y',w) \left[
{{\tilde \Psi_2(y,we^{2 \pi i})} \over {W(we^{2 \pi i})}} -
{{\tilde \Psi_2(y,w)} \over {W(w)}} \right] \nonumber \\
&& {_{s}S_{l}}(\theta,aw){_{s}S_{l}}(\theta',aw) e^{-iwt} 
dw\  .
\end{eqnarray}
(For simplicity we assume that the initial data has a considerable support 
only for $r$-values which are smaller than the observer's location.
This, of course, does not change the asymptotic {\it late}-time
behaviour).

Using the fact that $M(a,b,z)$ is a single-valued function and Eq. 13.1.10 of
\cite{Abram} (taking the $b \to k$ limit, where $k$ is an integer), 
one finds
\begin{equation}\label{Eq26}
\tilde \Psi_1(r,we^{2 \pi i})= \tilde \Psi_1(r,w)\  ,
\end{equation}
and
\begin{equation}\label{Eq27}
\tilde \Psi_2(r,we^{2 \pi i})=\tilde \Psi_2(r,w)+
{B \over A} { {(-1)^{2l} 2\pi i} \over {(2l+1)! \Gamma (-l+s -
2iwM )}} \tilde \Psi_1(r,w)\  .
\end{equation}
Using Eqs. (\ref{Eq26}) and (\ref{Eq27}) it is easy to see that
\begin{equation}\label{Eq28}
W(we^{2 \pi i})=W(w)\  .
\end{equation}
Thus, using Eqs. (\ref{Eq26}), (\ref{Eq27}) and (\ref{Eq28}), we obtain the
relation
\begin{equation}\label{Eq29}
{{\tilde \Psi_2(r,we^{2 \pi i})} \over {W(we^{2 \pi i})}} -
{{\tilde \Psi_2(r,w)} \over {W(w)}} ={B \over A}{{(-1)^{2l} 2 \pi i} \over 
{(2l+1)! \Gamma(-l +s -2iwM)}}
{{\tilde \Psi_1(r,w)} \over {W(w)}}\  .
\end{equation}
Since $W(w)$ is $r$-independent, we may use the large-$r$ asymptotic 
expansions of the confluent hypergeometric functions (given by Eqs. 13.5.1 and
13.5.2 in \cite{Abram}) in order to evaluate it. One finds
\begin{equation}\label{Eq30}
W(w)=i {AB (-1)^{l+1} (2l+1)! w^{-2l -1} \over
{\Gamma (l+s +1-2iwM) 2^{2l +1}}}\  .
\end{equation}
(Of coarse, using the $|z| \to 0$ limit of the confluent hypergeometric 
functions, we obtain the same result).
Thus, substituting Eqs. (\ref{Eq29}) and (\ref{Eq30}) in Eq. 
(\ref{Eq25}) one finds
\begin{eqnarray}\label{Eq31}
G^C(z,z';t)& =&  \sum\limits_{l=l_{0}}^{\infty} 
{{ (-1)^{l+1} 2^{2l}} \over {\pi A^{2} [(2l+1)!]^2}} 
\int_{0}^{-i \infty} {{\Gamma(l+s+1-2iwM)} 
\over {\Gamma(-l+s-2iwM)}} \nonumber \\
&& \tilde \Psi_1(y,w) 
\tilde \Psi_1(y',w){_{s}S_{l}}(\theta,aw){_{s}S_{l}}(\theta',aw)
w^{2l+1} e^{-iwt} dw\  .
\end{eqnarray}
As was explained, the late-time behaviour of the field should follow from the
{\it low}-frequency contribution to the Green's function.
It is clear that all the rotation-dependent terms 
of Eq. (\ref{Eq11}) can be neglected in the $aw \to 0$ limit.
Hence, one may replace the functions $_{s}S_{l}^{m}(\theta,aw)$
by the spin-weighted spherical harmonics $_{s}Y_{l}^{m}(\theta,\phi)$.
In addition, we use the approximation
\begin{equation}\label{Eq32}
{\Gamma(l+s+1-2iwM) \over \Gamma(-l+s-2iwM)} \simeq 
2iwM(-1)^{l-s+1}(l+s)!(l-s)!\  ,
\end{equation}
which is valid for $w \to 0$.
With these substitutions Eq. (\ref{Eq31}) becomes
\begin{eqnarray}\label{Eq33}
G^C(z,z';t)& =&  \sum\limits_{l=l_{0}}^{\infty} 
{{i M (-1)^{2l-s} 2^{2l+1} (l+s)! (l-s)!} \over {\pi A^{2} [(2l+1)!]^2}} 
{_{s}Y_{l}(\theta)} {_{s}Y_{l}^{*}(\theta')} \nonumber \\
&& \int_{0}^{-i \infty} \tilde \Psi_1(y,w) 
\tilde \Psi_1(y',w) w^{2l+2} e^{-iwt} dw\  .
\end{eqnarray}

\subsection{Asymptotic behaviour at timelike infinity}\label{Sec4B}

First, we consider the asymptotic
behaviour of the fields at {\it timelike infinity} $i_+$.
As was explained, the late-time behaviour of the field should follow from the
{\it low}-frequency contribution to the Green's function. Actually, it is easy 
to verify that the effective contribution to the integral in 
Eq. (\ref{Eq33}) should come from $|w|$=$O({1 \over t})$. 
Thus, in order to obtain the asymptotic
behaviour of the field at {\it timelike infinity} (where $y,y' \ll t$),
we may use the $|w|r \ll 1$ limit of $\tilde \Psi_1(r,w)$. Using Eq. 13.5.5
from \cite{Abram} one finds
\begin{equation}\label{Eq34}
\tilde \Psi_1(r,w) \simeq Ar^{l +1}\  .
\end{equation}
Substituting this in Eq. (\ref{Eq33}) we obtain
\begin{eqnarray}\label{Eq35}
G^C(z,z';t)&=& \sum\limits_{l=l_{0}}^{\infty} 
{{i M (-1)^{2l-s} 2^{2l+1} (l+s)!(l-s)!} \over {\pi [(2l+1)!]^2}}
{_{s}Y_{l}(\theta)} {_{s}Y_{l}^{*}(\theta')} (yy')^{l +1}
\nonumber \\ 
&& \int_{0}^{-i \infty} w^{2l +2}
e^{-iwt} dw\  ,
\end{eqnarray}
Performing the integration in Eq. (\ref{Eq35}), one finds
\begin{eqnarray}\label{Eq36}
G^C(z,z';t) &=& \sum\limits_{l=l_{0}}^{\infty}
{{M (-1)^{l-s+1} 2^{2l+1} (l+s)!(l-s)! (2l+2)!} \over {\pi [(2l+1)!]^2}}
\nonumber \\ 
&& {_{s}Y_{l}(\theta)} {_{s}Y_{l}^{*}(\theta')}
y'^{l+1} y^{l +1} t^{-(2l +3)}\  .
\end{eqnarray}

\subsection{Asymptotic behaviour at future null infinity}\label{Sec4C}

Next, we go on to consider the behaviour of the fields at
future null infinity $scri_+$ .
It is easy to verify that for this case
the effective frequencies contributing to the integral in Eq. (\ref{Eq33}) are
of order $O({1 \over u})$.
Thus, for $y-y' \ll t \ll 2y-y'$ one may use the
$|w|y' \ll 1$ asymptotic limit for $\tilde \Psi_1(y',w)$ and the $|w|y \gg 1$
($Imw < 0$) asymptotic limit of $\tilde \Psi_1(y,w)$. Thus,
\begin{equation}\label{Eq37}
\tilde \Psi_1(y',w) \simeq Ay'^{l +1}\  ,
\end{equation}
and
\begin{equation}\label{Eq38}
\tilde \Psi_1(y,w) \simeq Ae^{iwy} (2l+1)! {{e^{-i {\pi \over 2}
(l+s+1-2iwM)} (2w)^{-l-s-1+2iwM} y^{-s+2iwM}} \over
{\Gamma (l-s+1+2iwM)}}\  ,
\end{equation}
where we have used Eqs. 13.5.5 and 13.5.1 of \cite{Abram}, respectively.
Integrating Eq. (\ref{Eq33}) with the aid of 
Eqs. (\ref{Eq37}) and (\ref{Eq38}) one finds (for $v \gg u$)
\begin{eqnarray}\label{Eq39}
G^C(z,z';t) &=& \sum\limits_{l=l_{0}}^{\infty}
{{M (-1)^{l-s+1} 2^{l} (l+s)! (l-s+2)!(2l+2)} \over {\pi (l+2)!}}
\nonumber \\ 
&& {_{s}Y_{l}(\theta)} {_{s}Y_{l}^{*}(\theta')}
y'^{l+1} v^{-s} u^{-(l-s+2)}\  .
\end{eqnarray}
  
\subsection{Asymptotic behaviour along the black-hole outer horizon}
\label{Sec4D}

Finally, we consider the behaviour of the fields at the
black-hole outer-horizon $r_+$. While Eqs. (\ref{Eq23}) and (\ref{Eq24})
are (approximated) solutions to the wave-equation (\ref{Eq14}) in the
$r \gg M,|a|$ case, they do not represent the solution near the
horizon. As $y \to -\infty$ the wave-equation (\ref{Eq14}) can be
approximated by the equation \cite{Teukolsky2}
\begin{equation}\label{Eq40}
\tilde \Psi _{,yy} + \left[ k^{2}-{{2is(r_{+}-M)k} \over {2Mr_{+}}} -
{{s^{2}(r_{+}-M)^{2}} \over {(2Mr_{+})^{2}}} \right] \tilde \Psi =0\  .
\end{equation}
Thus, we take 
\begin{equation}\label{Eq41}
\tilde \Psi_1(y,w)=C(w) \Delta^{-s/2} e^{-iky}\  ,
\end{equation}
and we use Eq. (\ref{Eq37}) for $\tilde \Psi_1(y',w)$. In order to match
the $y \ll -M$ solution with the $y \gg M$ solution we assume that the
two solutions have the same temporal dependence (this assumption has
been proven to be very successful for neutral \cite{Gundlach} 
and charged \cite{HodPir1} perturbations on a spherically 
symmetric backgrounds).
In other words we take $C(w)$ to be $w$-independent.
In this case one should replace the roles of $y'$ and $y$ in
Eq. (\ref{Eq33}).
Using Eq. (\ref{Eq33}), we obtain
\begin{eqnarray}\label{Eq42}
G^C(z,z';t) &=& \sum\limits_{l=l_{0}}^{\infty}
\Gamma_{0} 
{{M (-1)^{l-s+1} 2^{2l+1} (l+s)!(l-s)! (2l+2)!} \over {\pi [(2l+1)!]^2}}
\nonumber \\ 
&& {_{s}Y_{l}(\theta)} {_{s}Y_{l}^{*}(\theta')}
\Delta^{-s/2} y'^{l+1} e^{imw_{+}y} v^{-(2l +3)}\  ,
\end{eqnarray}
where $\Gamma_{0}$ is a constant.

\subsection{Coupling of different multipoles}\label{Sec4E}

The time-evolution of a wave-field 
described by Eq. (\ref{Eq4}) is given by Eq. (\ref{Eq8}).
The coefficients $B_{1}(r,\theta)$ and $B_{2}(r,\theta)$ appearing in 
Eq. (\ref{Eq4}) depend explicitly on the angular variable $\theta$ through
the {\it rotation} of the black-hole (no such dependence exist in the
$a=0$ case). This angular dependence of the coefficients leads to 
an active interaction between different multipoles (characterized by
different values of $l$).
This coupling between different multipoles is physically caused by 
the {\it rotational} dragging of reference frames (due to the rotation
of  the black-hole).
Thus, even if the initial data is characterized by a pure multipole
$l$ (i.e., by a certain spin-weighted spherical 
harmonic $_{s}Y_{l}$) other multipoles would be 
generated dynamically during the evolution. In other words, 
the late-time evolution of a spin-$s$ field is dominated by the
lowest allowed multipole, i.e., by the $l=|s|$ multipole, regardless
of the angular dependence of the initial-data (provided that the
initial-data contains modes with $|m| \leq |s|$).
This phenomena of coupling between different multipoles has been observed
in numerical solutions of the Teukolsky 
equation \cite{Krivan1,Krivan2} (a similar phenomena 
is known in the case of rotating stars \cite{Kojima}).

\section{Summary}\label{Sec5}

We have studied the asymptotic late-time evolution of realistic {\it
  rotating} collapse. This was done by considering the 
asymptotic late-time solutions of Teukolsky's master 
equation which governs the evolution of gravitational, 
electromagnetic, neutrino and scalar (massless) perturbations fields 
on Kerr spacetimes.

Following the {\it no-hair conjecture} for rotating black-holes we 
have shown that the asymptotic solutions 
develop inverse {\it power-law} tails at timelike infinity, at
null infinity and along the black-hole outer horizon (where the
power-law behaviour is multiplied by an oscillatory term, caused by 
the dragging of reference frames at the event horizon).

The damping exponents, describing the fall-off of the asymptotic
solutions of Teukolsky's equation at timelike infinity and 
along the black-hole outer 
horizon are independent of the spin parameter of the field. 
However, we have shown that the damping exponents at future 
null infinity are {\it spin-dependent}.
Moreover, the asymptotic late-time solutions of Teukolsky's equation
at all the three asymptotic regions have a spatial dependence on the
spin parameter of the field.

The damping exponents are {\it independent} of the rotation parameter
$a$ of the black-hole (or star). 
However, the {\it rotational} dragging of reference frames, 
caused by the rotation of the black-hole has two important
effects (which are not found in the non rotating case) on the
asymptotic late-time evolution of {\it rotating} collapse:
The power-law tail along the black-hole outer horizon 
is multiplied by an {\it oscillatory} term. 
In addition, the rotation leads to an active {\it coupling} 
of different multipoles. Hence, the late-time evolution 
of a spin-$s$ field is dominated by the lowest allowed 
multipole, i.e., by the $l=|s|$ multipole.

\bigskip
\noindent
{\bf ACKNOWLEDGMENTS}
\bigskip

This research was supported by a grant from the Israel Science Foundation.

\end{document}